# On the Correct Use of Statistical Tests:
## Reply to «Lies, damned lies and statistics (in Geology)»

D. Sornette[1-3], V.F. Pisarenko[4]

[1]ETH Zurich, D-MTEC, Kreuzplatz 5
CH-8032 Zurich, Switzerland
[1]ETH Zurich, Department of Earth Sciences
[3]Institute of Geophysics and Planetary Physics
University of California, Los Angeles, California 90095
[4]International Institute of Earthquake Prediction Theory
and Mathematical Geophysics
Russian Ac. Sci., Profsoyuznaya 84/32, Moscow 117997, Russia
Emails: dsornette@ethz.ch and pisarenko@yasenevo.ru

**Abstract**: *In a Forum published in EOS Transactions AGU (2009) entitled "Lies, damned lies and statistics (in Geology)", Vermeesch (2009) claims that "statistical significant is not the same as geological significant", in other words, statistical tests may be misleading. In complete contradiction, we affirm that statistical tests are always informative. We detail the several mistakes of Vermeesch in his initial paper and in his comments to our reply. The present text is developed in the hope that it can serve as an illuminating pedagogical exercise for students and lecturers to learn more about the subtleties, richness and power of the science of statistics.*

**1-Reply to Vermeesch (2009)**

In a recent Forum in EOS entitled «Lies, Damned Lies and Statistics (in Geology)», Vermeesch (2009) applied the standard Chi-square test to a global catalog of earthquakes (USGS, http://earthquake.usgs.gov; *118,415 events, $4.0 \leq m \leq 9.0$, 1999-2009*) and revealed that the distribution of global seismicity over weekdays is not uniform. The chi-square sum $S(D)$ is *94*, which corresponds to an extremely small *p*-value = *$4.5 \cdot 10^{-18}$* for the null hypothesis that «the occurrence of earthquakes does not depend on the day of the week». This makes the null hypothesis of the uniform distribution of seismicity over week days absolutely unacceptable.

Then, Vermeesch applied to the catalog the following operation: "Using the same proportion of earthquake occurrences but reducing the sample size by a factor *10* results in a *10* times smaller chi-square value *$(S(D) = 9.4)$,* corresponding to a *p*-value of *0.15*, which is greater than *0.05* and fails to reject the null hypothesis. In conclusion, the strong dependence of *p*-values on sample size makes them uninterpretable". Vermeesch concluded that «statistical significant is not the same as geological significant».

In complete contradiction, we affirm that statistical tests, if they are used properly, are always informative.

The conclusions by Vermeesch are erroneous, and the error consists in the inadmissibility of the operation consisting of simultaneously dividing the total sample size and the earthquake occurrences in each weekday by *10* (or any other factor). The chi-square sum is made of normalized squared deviations of observed frequencies from theoretical ones (all equal to *1/7* under the null hypothesis of uniform seismicity over the 7 weekdays). Dividing the sample size by a factor of *10* results in decreasing the mean values of the squared deviations by $10^2$, and in decreasing the variances of the deviations by *10*. Thus, their ratio is decreased by 10, as found by Vermeesch. It is essential to realize that such an operation does not correspond to an apparently harmless reduction of the sample size from *118,415* to *11,842*. In particular, the ten-fold reduction of the chi-square sum *S(D)* is inadmissible. Indeed, under the condition that the sample is large enough and the observations are independent and identically distributed, then a standard result states that the *p*-value practically does not depend on sample size, since the asymptotic chi-square distribution is valid for any sufficiently large sample size satisfying the condition of a minimum number of observations per bin (usually 10). In a nutshell, the core of the error of Vermeesch is that he created a new data set by taking 10% of the data in each bin (day of the week). This created a biased sample through the forced equal reduction of data in each bin. Instead, Vermeesch should have taken 10% of the original data set and then grouped it into 7 bins again. When performing this correct procedure, we find a *p*-value for the reduced sample size of about $10^{-6}$, instead of *0.15* in Vermeesch's procedure. Thus, decreasing the sample size by 10 still rejects the null hypothesis that the occurrence of earthquakes does not depend on the day of the week.

In order to interpret the cause of the rejection of the null hypothesis, both for the initial sample and for the (correctly) reduced one, we need to recall that the chi-square test is based on the asymptotical distribution (as sample size *n* tends to infinity) of the normalized, squared sum of deviations of the observed frequencies from theoretical ones. The chi-square test should be applied to ***independent, identically distributed random observations***. In addition, the distribution of observations over bins should satisfy the condition that each bin contains not less than *8-10* observations. As long as the last condition is satisfied, the *p*-value almost does not depend on the sample size (a possible weak dependence asymptotically vanishes). Since geological causes of a heterogeneous distribution of earthquakes over different weekdays seem improbable, one can assume that some conditions of applicability of the chi-square test are violated in the case of the earthquake catalog studied by Vermeesch (2009). We can enumerate at least five reasons of such violation:

- aftershocks;
- so-called "swarms" of weak shocks (of vague tectonic nature);
- artificial seismic events (quarry blasts; fluid-induce seismicity and so on, see e.g. [Goldbach, 2009];
- lower background noise on week-ends
- catalog incompleteness.

Of course, there can exist other reasons of non-stationarity or interdependence of events, preventing a justified use of the chi-square test.

We are going to remove aftershocks and other possible clusters from the catalog in order to test the null hypothesis for the remaining set of main shocks. This is done to obtain a catalog of earthquakes, which obeys better one of the conditions for the application of the chi-square test, namely the independence between events. For this, a standard procedure in seismology is to

«decluster» earthquake catalogs by removing as much as possible the aftershocks. Here, we apply the "aftershock cleaning" method described in details in [Pisarenko et al., 2008]. We found *80616* aftershocks (constituting *68% of the total number of events*), and *118415 – 80616 = 37799* main events. The main events are distributed as follows in each weekday: Mon 5135; Tue 5423; Wed 5338; Thu 5615; *Fri 5218; Sat 5485;* Sun 5585. The chi-square sum $S$ for the main shocks is

$$S = \sum_{k=1}^{7} [n_k - (1/7) \cdot n]^2 / [(1/7) \cdot n] , \qquad (1)$$

where $n_k$ is the number of events in day $k$ of the week and $n$ is the total number of events ($n = n_1 + n_2 + n_3 + n_4 + n_5 + n_6 + n_7$). We find $S = 36.19$, with $p = 2.5 \cdot 10^{-6}$. Although the *p*-value for the main shocks increases a lot, as compared with the *p*-value including the aftershocks, its small value still leads to reject the null hypothesis of a uniform distribution of events over the weekdays. While the "aftershock cleaning" method removes an essential portion of the aftershocks, no declustering method is perfect. In addition, we have not addressed the possible effect of "swarms" of weak shocks and of explosions. These effects refer to weak events, and we have little information on the nature of these events (if any). We know only that the Gutenberg-Richer law should be fulfilled (at least in the range of moderate events) in order for the catalog of earthquakes to be a complete representative one. For this purpose, we are going to truncate the catalog to remove weak events which are below a threshold that is higher than the threshold *m=4.0* given by the catalog. This gets rid of possible artificial shocks and allows us to analyze the remaining events, which should obey even better the conditions for the application of the chi-square test.

Figure 1 shows in the inset that the Gutenberg-Richer law for the distribution of earthquake magnitudes is fulfilled rather satisfactorily. Some deviations can be seen at lower magnitudes *m < 5* and above *m = 7.5*. These features are documented extensively in the seismological literature. The histogram of earthquake magnitudes in Fig. 1 shows that the monotonic decrease of the distribution begins at *4.5*, but the threshold of completeness should be somewhat larger, approximately for $m \geq 5.0$. Many seismologists believe that the completeness threshold for the Harvard global catalog is about *m=5.5* (since 1987) and *m=5.75* (since 1977), see e.g. [Molchan et al., 1996]. Being a bit less restrictive and selecting only earthquakes with $m \geq 5.0$ but without removing the aftershocks keeps *16308* events. The events with $m \geq 5.0$ are distributed as follows in each weekday: *Mon 2374; Tue 2511; Wed 2291; Thu 2497; Fri 2153; Sat 2282; Sun 2360.* The chi-square sum $S$ given by expression (1) is $S = 42.38$, with $p = 1.55 \cdot 10^{-7}$.

Combining the declustering "aftershock cleaning" method in order to obtain approximately independent events and working with a more complete catalog with only events of magnitude $m \geq 5.0$ leads us to identify *10672* aftershocks *(65%),* and *16308 – 10672 = 5636* main events with $m \geq 5.0$. The main events with $m \geq 5.0$ are distributed as follows in each weekday: *Mon 780; Tue 847;* Wed 793; *Thu 831;* Fri 785; *Sat 821; Sun 779.* The chi-square sum $S$ given by expression (1) is $S = 5.64$, with $p = 0.46$. Thus, the hypothesis that "the occurrence of earthquakes does not depend on the day of the week" is not rejected for large main shocks.

We can thus affirm that the main earthquake shocks with $m \geq 5.0$ are distributed uniformly over the seven weekdays, as expected from "seismological intuition." We have obtained this result correcting Vermeesch (2009)'s conclusion by taking into account the two most important properties of earthquake catalogs: the presence of numerous aftershocks and the problem of catalog incompleteness [Kagan, 2003].

The smaller earthquakes (main shocks included) might be distributed unevenly during the week, but elucidating the origin and nature of this phenomenon require much more additional information about the explosions, swarms of weak events and so on [Atef et al., 2009].

When properly used and interpreted, statistical tests are *always* revealing useful information. Ridiculously small *p*-values as found by Vermeesch (2009) should lead to questioning one by one all the assumptions which underpin the used statistical test. Geological hypotheses ought to include in the null hypothesis that earthquakes are correlated and that there are data uncertainties. A full model is needed to correctly calculate appropriate significance levels. Such approach is now undertaken since many years by different professional statistical seismologists [Keilis-Borok and Soloviev, 2003; Schorlemmer et al., 2007; 2009].

Mathematics is not wrong, only its incorrect interpretation may lead to confusion and paradoxes.

**2-Reply to Vermeesch (2011)**

**Chi-square test versus Neyman-Pearson testing**

Vermeesch (2009)'s main message was expressed in its title: the statistics (in geology) is something like lies or even damned lies. In his reply of 2001 ("Statistical Significance Does Not Equal Geological Significance"), Vermeesch shows no desire to withdraw his words, appearing to hold on his views. Curiously enough, in order to motivate his argument, Vermeesch uses some statistics and even a citation of the great statistician John Tukey (whom he, in accordance with his opinion, should consider perhaps as a damned liar). As an illustration of his message, Vermeesch takes an example the seismology), namely the distribution of seismic shocks in a catalog of the US Geological Survay, 1999-2009, (118415 events) over week days. As we explained in the first section, the gross error (from a statistical view point) is to reduce the sample size (n = 118415) by a factor 10, keeping the same proportions of earthquake occurrences over weekdays. Such operation is erroneous and absolutely inadmissible.

In his reply "Statistical Significance Does Not Equal Geological Significance", Vermeesch (2011) makes a second gross error. He confuses the standard Chi-square test with the Neyman-Pearson testing. The first one is a typical statistical ***test of significance*** (see e.g. Harald Cramer "Mathematical Methods of Statistics", Princeton Univ. Press, Princeton, N.-Y., 1940, Chapter 39). The test of significance works only with a null hypothesis. In our case, it is the hypothesis assuming a uniform distribution of earthquakes throughout the week. No alternative hypothesis is considered in the significance test. One either rejects the null hypothesis, if the p-value is small, say, less than 0.10, or accepts it, if p > 0.10. This Chi-square test was specifically considered in "Lies, Damned Lies, and Statistics (in Geology)". Vermeesch got $p = 4.5 \cdot 10^{-18}$ and rejected the null hypothesis. Then, having got p = 0.15 as result of the first erroneous operation mentioned above, Vermeesch claims that p-values are unstable, and essentially useless.

In contrast, the Neyman-Pearson approach assumes an alternative hypothesis (or a family of such alternatives). No such alternative was mentioned in "Lies, Damned Lies, and Statistics (in Geology)" (Vermeesch, 2009). Then, Vermeesch (2011) seems to adopt a new strategy of confusing the reader to take the attention away from his first erroneous point by discussing the Neyman-Pearson test framework. In this case, an alternative hypotheses should be introduced. Thus, Vermeesch somehow introduces (not explicitly, not clearly) alternative hypotheses. One cannot say much about these alternatives. One can only guess about their existence by the fact that, in Table 1 of (Vermeesch, 2011), the results are shown for a non-central Chi-square distribution (NCSD). In the standard Chi-square test, there should not be any NCSD. There was no word about NCSD in "Lies, Damned Lies, and Statistics (in Geology)" (Vermeesch, 2009). In (Vermeesch, 2011), it appears.

Table 1 of (Vermeesch, 2011) is contradictory and hardly understandable . On the one hand, it refers to calculations under an alternative hypothesis (see words "Power Calculation" in the title and the words on the non-central Chi-square distribution in the caption). On the other hand, we see words "expected p Value" in the third column of Table, which means that it deals with a null hypothesis (see the author comment in the second paragraph: "by definition, the p value is the probability *under the null hypothesis*, of obtaining the test statistic at least as extreme as the one observed"). Table 1 illustrates an obvious result: the smaller is a deviation from the null hypothesis, the more observations are needed to detect it. Or, in other words, the error of the second type $\beta$ increases when sample size n decreases **under the alternative hypothesis**. Thus, *under the alternative hypothesis*, the probability of exceeding observed statistic (*it is not the p-value, since the p-value refers to the null hypothesis!*) does depend on sample size. But **under the null hypothesis** there is no such dependence. Vermeersch says: "Table 1 strongly contradicts the claim by Sornette and Pisarenko (2011) that `the p value should not depend on sample size' ". Vermeersch misuses our words. We said the following: "under the conditions that the sample is large enough, that observations are independent and identically distributed, and *the zero hypothesis is true*, the p value should not depend essentially on sample size, as long as there are at least 10 observations per bin". This statement is absolutely true. Indeed, the p-value under the null hypothesis is distributed uniformly on the interval (0, 1). This distribution is well determined as soon as the sample size n becomes large enough and remains steady for all larger n. This is the result of the classical theorem about the Chi-square test proved by K. Pearson (1900). So, once more Vermeesch is mistaken.

Unfortunately, most of the other papers in the forum discussing (Vermeersch, 2009) start by discussing the second subject suggested by Vermeesch: "what can be the possible reasons for the unevenly distribution of earthquakes throughout the week established by the Chi-square test?" There are some informative and interesting considerations on this subject in the discussion, but they carry away from the main question: Whether or not statistics are damned lies? Or did Vermeesch perform gross errors?

**Acknowledgements**: We are thankful to Pieter Vermeesch for providing the earthquake catalog in question for exact one to one comparisons and to Max Werner for useful discussions.

**Bibliography**

Atef, A.H., Liu, K.H., and S. S. Gao (2009), Apparent Weekly and Daily Earthquake Periodicities in the Western United States, Bulletin of the Seismological Society of America 99 (4), 2273-2279.

Kagan, Y.Y. (2003), Accuracy of modern global earthquake catalogs, Phys. Earth Planet. Inter. 135 (2-3), 173-209.

Kagan, Y.Y. (2004) Short-Term Properties of Earthquake Catalogs and Models of Earthquake Source, Bulletin of the Seismological Society of America 94 (4), 1207-1228.

Keilis-Borok, V. I., and Soloviev, A. A. eds. (2003), Nonlinear Dynamics of the Lithosphere and Earthquake Prediction, Springer-Verlag, Heidelberg.

Molchan et al. (1996), Seismic risk oriented multiscale seismicity model: Italy, Computational Seismology, Iss. 28, pp. 193-224 (in Russian). See as well the English translation: Computational Seismology and Geodynamics, D.C.: American Geophysical Union, 1999, 200p.

Pearson, K. (1900) On the criterion that a given system of deviations from the probable in the case of a correlated system of variables is such that it can be reasonably supposed to have arisen from random sampling, Phil.Mag. 50, 157-175.

Pisarenko V.F., Sornette A., Sornette D., and M.V. Rodkin (2008) New Approach to the Characterization of Mmax and of the Tail of the Distribution of Earthquake Magnitudes, Pure and Applied Geophysics 165, 847-888.

Schorlemmer, D., M. C. Gerstenberger, S. Wiemer, D. D. Jackson, and D. A. Rhoades (2007), Earthquake Likelihood Model Testing, Seismological Research Letters 78 (1), 17-29.

Schorlemmer, D., J. D. Zechar, M. Werner, D. D. Jackson, E. H. Field, T. H. Jordan, and the RELM Working Group (2009), First results of the Regional Earthquake Likelihood Models Experiment, Pure and Applied Geophysics, submitted
(e-print at http://www.cseptesting.org/sites/default/files/zechar2007.pdf)

Sornette, D. and V. Pisarenko (2011), On the Correct Use of Statistical Tests: Reply to "Lies, damned lies and statistics (in Geology)", Eos Transactions AGU, Vol. 92, No. 8, page 64, 22 February 2011.

Vermeesch, P. (2009), Lies, Damned lies and statistics (in geology), Eos Transactions AGU 90 (47), 24 Nov 2009, p. 443.

Vermeesch, P. (2011), Statistical significance does not equal geological significance (reply to comments on "Lies, damned lies, and statistics (in Geology)"), Eos Transactions AGU 92 (8), 22 February 2011, p. 66.

Goldbach, O.D. (2009), Flood-induced seismicity in mines, 11-th SAGA Biennial Techn. Meeting and Exhibition, Swaziland, 16-18 Sept. 2009, pp.391-401. See in particular Fig.15 of this paper, showing an uneven distribution of weak seismic shocks over the week days, due to flood induced seismicity.

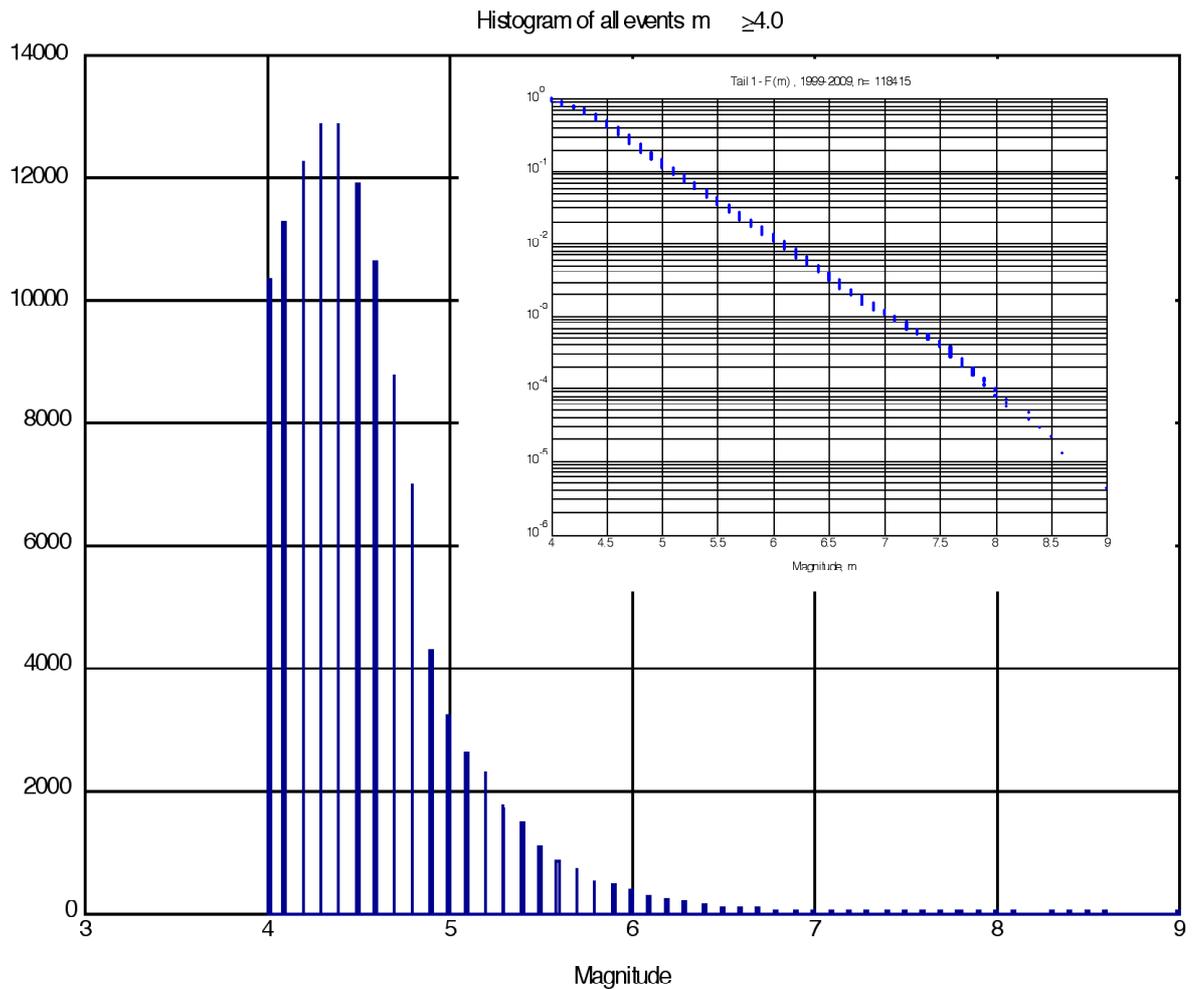

Fig.1: Histogram approximation of the distribution of earthquake magnitudes for all 118'415 events of magnitude 4 or greater and occurring between Friday 1st January 1999 and Thursday, 1 January 2009 (USGS, http://earthquake.usgs.gov). Inset: Decimal logarithm of the empirical complementary cumulative distribution function $1 - F(m)$ of earthquake magnitude.